\documentclass[prb,showpacs,preprintnumbers,twocolumn]{revtex4}
\usepackage{graphicx}
\newcommand{\cupzclo}{Cu(pz)$_2$(ClO$_4$)$_2$}
\pdfoutput=1
\begin{document}
\bibliographystyle{apsrev}
\title{The Two-Dimensional Square-Lattice S=1/2 Antiferromagnet \cupzclo}

\author{N.~Tsyrulin$^{1,2}$,
F.~Xiao$^{3}$, A.~Schneidewind$^{4,5}$, P.~Link$^{4}$,
H.~M.~R{\o}nnow$^{6}$, J.~Gavilano$^{2}$, C.~P.~Landee$^{3}$,
M.~M.~Turnbull$^{7}$ and M.~Kenzelmann$^{1,8}$}

\affiliation{(1) Laboratory for Solid State Physics, ETH Zurich,
CH-8093 Zurich, Switzerland \\(2) Laboratory for Neutron Scattering,
ETH Zurich $\&$ Paul Scherrer Institute, CH-5232 Villigen,
Switzerland \\(3) Department of Physics, Clark University,
Worcester, Massachusetts 01610, USA \\(4) Forschungsneutronenquelle
Heinz Meier-Leibnitz (FRM II), D-85747 Garching, Germany \\(5)
Institut f\"{u}r Festk\"{o}rperphysik, TU Dresden, D-01062 Dresden,
Germany \\(6) Laboratory for Quantum Magnetism,  \'{E}cole
Polytechnique F\'{e}d\'{e}rale de Lausanne (EPFL), CH-1015 Lausanne,
Switzerland
\\(7) Carlson School of Chemistry and Biochemistry, Clark
University, Worcester, Massachusetts 01610, USA \\(8) Laboratory for
Developments and Methods, Paul Scherrer Institute, CH-5232 Villigen,
Switzerland}
\begin{abstract}
We present an experimental study of the two-dimensional S=1/2
square-lattice antiferromagnet \cupzclo \ (pz denotes pyrazine -
$\rm C_4H_4N_2$) using specific heat measurements, neutron
diffraction and cold-neutron spectroscopy. The magnetic field
dependence of the magnetic ordering temperature was determined from
specific heat measurements for fields perpendicular and parallel to
the square-lattice planes, showing identical field-temperature phase
diagrams. This suggest that spin anisotropies in \cupzclo \ are
small. The ordered antiferromagnetic structure is a collinear
arrangement with the magnetic moments along either the
crystallographic b- or c-axis. The estimated ordered magnetic moment
at zero field is $\rm m_0=0.47(5)\;\mathrm{\mu_B}$ and thus much
smaller than the available single-ion magnetic moment. This is
evidence for strong quantum fluctuations in the ordered magnetic
phase of \cupzclo. Magnetic fields applied perpendicular to the
square-lattice planes lead to an increase of the
antiferromagnetically ordered moment to $\rm
m_0=0.93(5)\;\mathrm{\mu_B}$ at $\rm \mu_0H=13.5\;\mathrm{T}$ -
evidence that magnetic fields quench quantum fluctuations. Neutron
spectroscopy reveals the presence of a gapped spin excitations at
the antiferromagnetic zone center, and it can be explained with a
slightly anisotropic nearest neighbor exchange coupling described by
$\rm J_1^{xy}=1.563(13)\;\mathrm{meV}$ and $\rm
J_1^z=0.9979(2)J_1^{xy}$.
\end{abstract}
\maketitle
\section{Introduction.}
Low dimensional quantum magnets are of great fundamental interest.
Unlike three-dimensional magnets, they support strong quantum
fluctuations which can result in novel quantum excitations and novel
ground states. Case in point is the antiferromagnetic S=1 chain
whose ground state features hidden quantum order and is separated by
a finite energy from excited states \cite{Haldane, Buyers}. In
contrast, antiferromagnetic S=1/2 Heisenberg chains are gapless and
feature fractionalized spin excitations as the hallmark of quantum
criticality \cite{Tennant}.
\par
Increasing the dimensionality of a quantum magnet from one to two
dimensions generally reduces effects of quantum fluctuations. The
ground state of S=1/2 square lattice Heisenberg antiferromagnet (AF)
adopts N\'{e}el long-range order at zero temperature. Nevertheless,
strong quantum fluctuations arising from geometrical frustration may
destroy long-range order in two dimensions.
\par
Numerical studies of the two-dimensional (2D) S=1/2 Heisenberg AF on
a square lattice using quantum Monte Carlo, exact diagonalization,
coupled cluster as well as series expansion calculations reveal a
quantum renormalization of the one-magnon energy in the entire
Brillouin zone and the existence of a magnetic continuum at higher
energies \cite{Singh, Sandvik, Zheng, Ho, igarashi, singh}. In
recent years, quantum renormalization effects at zero field have
been studied using neutron scattering in a number of good
realizations of S=1/2 square-lattice Heisenberg AFs \cite{clarke2,
ronnow, Harris, Christensen, lumsden, McMorrow}. The addition of
antiferromagnetic next-nearest neighbor (NNN) interactions
destabilizes the antiferromagnetic ground state and increases
quantum fluctuations: according to the $\rm J_1-J_2$ model
\cite{Chandra, Chandra2, Read, Viana}, where $\rm J_1$ and $\rm J_2$
are the nearest neighbor (NN) and the NNN exchange interactions,
respectively, different ground states are stabilized as a function
of $\rm J_2/J_1$. A possible spin-liquid phase appears to be the
ground state for $\rm 0.38<J_2/J_1<0.6$ and collinear order was
found for $\rm J_2/J_1>0.6$. Our previous study \cite{tsyrulin} of
\cupzclo \ has shown that even a small $\rm J_2/J_1\simeq0.02$ ratio
enhances quantum fluctuations drastically, leading to a strong
magnetic continuum at the antiferromagnetic zone boundary and the
inversion of the zone boundary dispersion in magnetic fields.
\par
Here we present an experimental investigation of the 2D
organo-metallic AF \cupzclo, a good realization of the weakly
frustrated $\rm J_2/J_1\simeq0.02$ quantum AF on a square lattice
with $\rm J_1\sim1.56\;\mathrm{meV}$. Due to the small energy scale
of the dominant exchange interaction, magnetic fields available for
macroscopic measurements and neutron scattering allow the
experimental investigation of this interesting model system for
magnetic fields up to about one third of the saturation field
strength. We combine specific heat, neutron diffraction and neutron
spectroscopy to determine the spin Hamiltonian and the key magnetic
properties of this model material. Specific heat measurements show
that the magnetic properties are nearly identical for fields applied
parallel and perpendicular to the square-lattice plane. This shows
that spin anisotropies are small in contrast to spatial
anisotropies, and that it is sufficient to perform microscopic
measurement for just one field direction. Our microscopic neutron
measurements, on the other hand, provide information on the spin
Hamiltonian that explain the nearly identical HT phase diagrams for
the two field directions. Specific heat and neutron measurements of
\cupzclo \ thus ideally complement each other.
\begin{figure}
\begin{center}
  \includegraphics*[width=8.5cm,bbllx=0,bblly=0.65,bburx=1,bbury=1.2,angle=0,clip=]{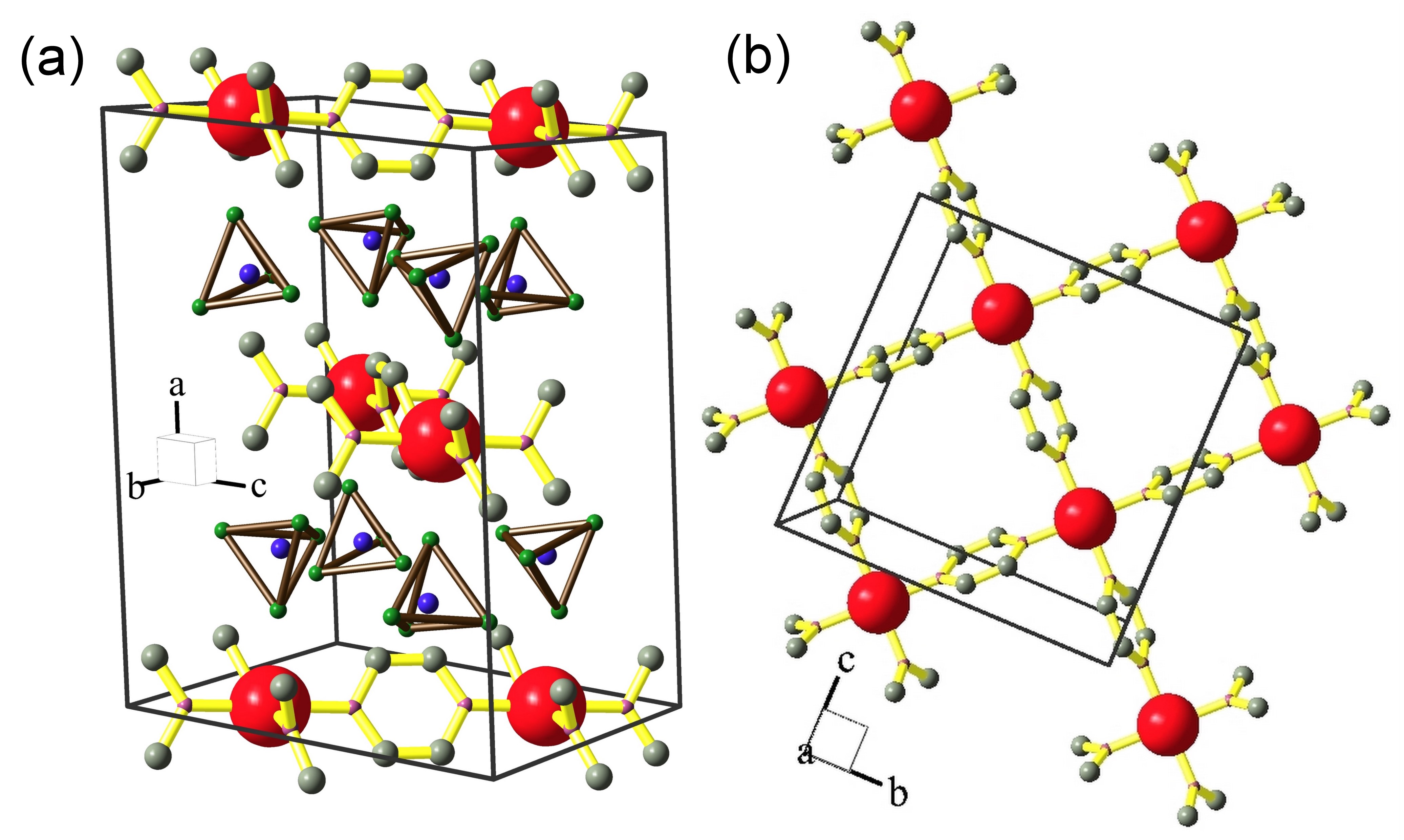}
  \caption{(Color online) (a) Three-dimensional view of the crystal structure of \cupzclo. The $\rm Cu^{2+}$ ions
are shown as big spheres. The $\rm ClO_4$ tetrahedra are located
between the copper layers and pyrazine molecules link ${\rm
Cu^{2+}}$ in $bc$-plane. The D atoms are not shown for simplicity.
(b) The projection of the crystal structure on the $bc$-plane shows
the ${\rm Cu^{2+}}$ square-lattice structure. The square lattice are
shifted by (0,0.5,0) from one layer to the next.}
  \label{structure}
\end{center}
\end{figure}
\par
Deuterated copper pyrazine perchlorate \cupzclo \ crystallizes in a
monoclinic crystal structure described by space group C2/c, with
lattice parameters $\rm a=14.045(5){\AA}$, $\rm b=9.759(3){\AA}$,
$\rm c=9.800(3){\AA}$ and $\rm \beta=96.491(4)^{\circ}$\ \cite{Wo}.
The crystal structure is shown in Fig.~\ref{structure}. The ${\rm
Cu^{2+}}$ ions occupy 4e Wyckoff positions and pyrazine ligands link
magnetic $\rm Cu^{2+}$ ions into square-lattice planes lying in the
crystallographic $bc$-plane. The ${\rm Cu^{2+}}$-${\rm Cu^{2+}}$ NN
distances in the $bc$-plane are identical and equal to 6.92$\rm {\AA}$ \cite{Wo}. The two fold rotation axis (0, y, 1/2) and the mirror
plane parallel to the $ac$-plane ensure that all NN exchange
interactions between $\rm Cu^{2+}$ are identical. Tetrahedra of
ClO$_4$ located between the planes (Fig.~\ref{structure}a) provide
good spatial isolation of $\rm Cu^{2+}$ ions and substantially
decrease the interlayer interactions. Thus, perfect square-lattices
of copper ions with a superexchange path mediated by pyrazine
molecules are formed in the $bc$-plane (Fig.~\ref{structure}b).
\par
The magnetic susceptibility shows good agreement with that of the 2D
S=1/2 Heisenberg AF with an exchange interaction strength of $\rm
J_1 = 1.53(3)\;\mathrm{meV}$. Small interlayer interactions result
in a long range antiferromagnetic order below $\rm
T_N=4.21(1)\;\mathrm{K}$. The ratio of interlayer exchange, $\rm J_{\perp}$,
to the dominant intralayer exchange strength, $\rm J_1$,  was estimated as $\rm
J_{\perp}/J_1=8.8\cdot10^{-4}$ \ \cite{Wo, Lancaster, Xiao}.
\par
\begin{figure}
\begin{center}
  \includegraphics*[width=7.5cm,bbllx=0,bblly=0,bburx=1,bbury=1.2,angle=0,clip=]{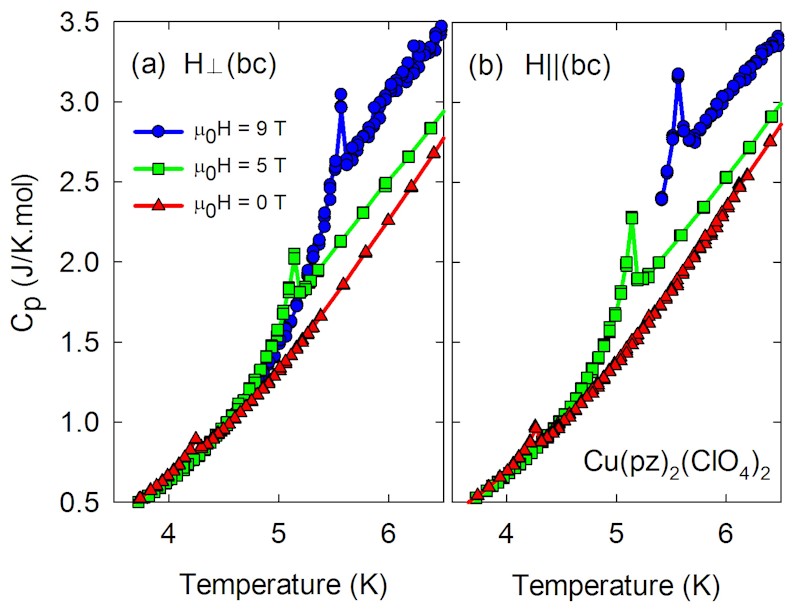}
  \caption{(Color online) Specific heat of \cupzclo \ as a function of temperature for different magnetic field strengths. The measurements
  for magnetic fields parallel and perpendicular to the $bc$-plane are shown in (a) and (b), respectively.
  For convenience the graphs are shown as lines connecting the data points and without error bars.}
  \label{Cp}
\end{center}
\end{figure}
\section{Experimental details.}
In order to obtain the field-temperature (HT) phase diagram of
\cupzclo \ we measured the specific heat as a function of
temperature for different magnetic field strengths using the
Physical Property Measurement System by Quantum Design. A single
crystal of deuterated \cupzclo \ with mass $\rm m=13\;\mathrm{mg}$
was fixed on a sapphire chip calorimeter with Apiezon-N grease. The
measurements were done using the relaxation technique which consists
of the application of a heat pulse to a sample and the subsequent
tracking the induced temperature change. The specific heat was
obtained in the range from $\rm T=2\;\mathrm{K}$ to $\rm
T=30\;\mathrm{K}$ in magnetic fields of up to $\rm
\mu_0H=9\;\mathrm{T}$ applied parallel and perpendicular to the
copper square-lattice planes. The measurements were done with the
steps of $\rm \Delta T_1=0.05\;\mathrm{K}$, $\rm \Delta
T_2=0.2\;\mathrm{K}$ and $\rm \Delta T_3=1\;\mathrm{K}$ in the
temperature ranges $\rm T_1=2 - 6\;\mathrm{K}$, $\rm T_2=6 -
8\;\mathrm{K}$ and $\rm T_3=8 - 30\;\mathrm{K}$, respectively. Care
was taken to apply a small heat pulse of $\rm 0.1\%$ of the
temperature step $\rm \Delta T$ and each measurement was repeated
three times to increase accuracy. Specific heat of Apiezon-N grease
without \cupzclo \ crystal was measured in the entire temperature
range separately and subtracted as a background from the total
specific heat of the sample and grease.
\begin{figure}
\begin{center}
  \includegraphics*[width=8cm,bbllx=0,bblly=0,bburx=1,bbury=0.85]{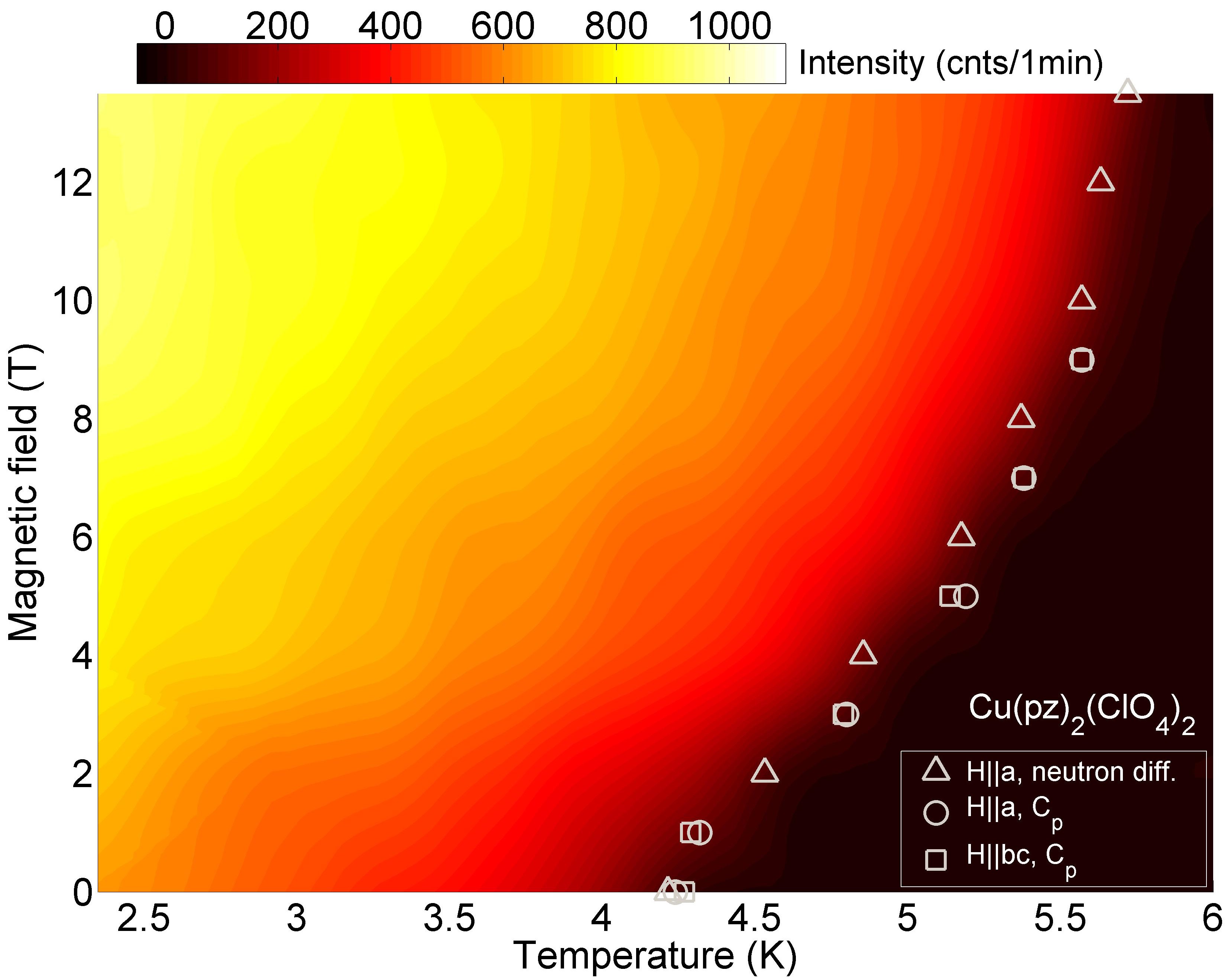}
  \caption{(Color online) Neutron scattering peak intensity of $\textbf{Q}=(0, 1, 0)$ as a function of
  temperature and magnetic field, applied perpendicular to the square-lattice
  planes. The results obtained by specific heat measurements in magnetic field applied parallel and perpendicular
  to the Cu planes are shown by squares and circles, respectively. The neutron data, measured for magnetic fields perpendicular to the
  copper planes, are shown by triangles.}
  \label{ht}
\end{center}
\end{figure}
\par
The HT phase diagram and the ordered magnetic structure were studied
by neutron diffraction using cold-neutron three-axis spectrometer
RITA2 at the Paul Scherrer Institute, Villigen, Switzerland. A
crystal with dimensions $7\times7\times1.5\;\mathrm{mm}$ and mass of
$\rm m=85\;\mathrm{mg}$ was wrapped into aluminum foil, fixed with
wires on a sample holder and aligned with its reciprocal [0, k, l]
plane in the horizontal scattering plane of the neutron
spectrometer. Data were collected at $\rm T=2.3\;\mathrm{K}$ and
$\rm T=10\;\mathrm{K}$ in magnetic fields up to $\rm
\mu_0H=13.5\;\mathrm{T}$ applied nearly perpendicular to the [0, k,
l] plane using an Oxford cryomagnet. Measurements were performed
with the pyrolytic graphite (PG) (002) Bragg reflection as a
monochromator. A cooled Be filter was installed before the analyzer
to suppress higher order neutron contamination for the final energy
$\rm E_f=5\;\mathrm{meV}$. We also used an experimental setup
without Be filter, which allowed to use the second order neutrons
from the monochromator with $\rm E_i=20\;\mathrm{meV}$, thus
allowing to access to reflections at high wave-vector transfers.
\par
The spin dynamics in the antiferromagnetically ordered phase was
measured using the cold-neutron three-axis spectrometer PANDA at
FRM-2, Garching, Germany. Two single crystals with a total mass of
$\rm m=1\;\mathrm{g}$ were wrapped into aluminum foil, fixed on a
sample holder with wires and co-aligned in an array with a final
mosaic spread of $1^{\circ}$. Reciprocal [0, k, l] plane of the
sample was aligned with the horizontal scattering plane of the
neutron spectrometer. These measurements were performed in zero
magnetic field and at temperature $\rm T=1.42\;\mathrm{K}$ using a
$\rm ^4He$ cryostat generally referred to as an Orange cryostat. The
final energy was either set to $\rm E_f=4.66\;\mathrm{meV}$ or $\rm
E_f=2.81\;\mathrm{meV}$ using a PG(002) analyzer. Data were
collected using a PG(002) monochromator and cooled Be filter
installed before the analyzer.
\begin{figure}
\begin{center}
  \includegraphics*[width=8.5cm,bbllx=0,bblly=0,bburx=1,bbury=0.725,angle=0,clip=]{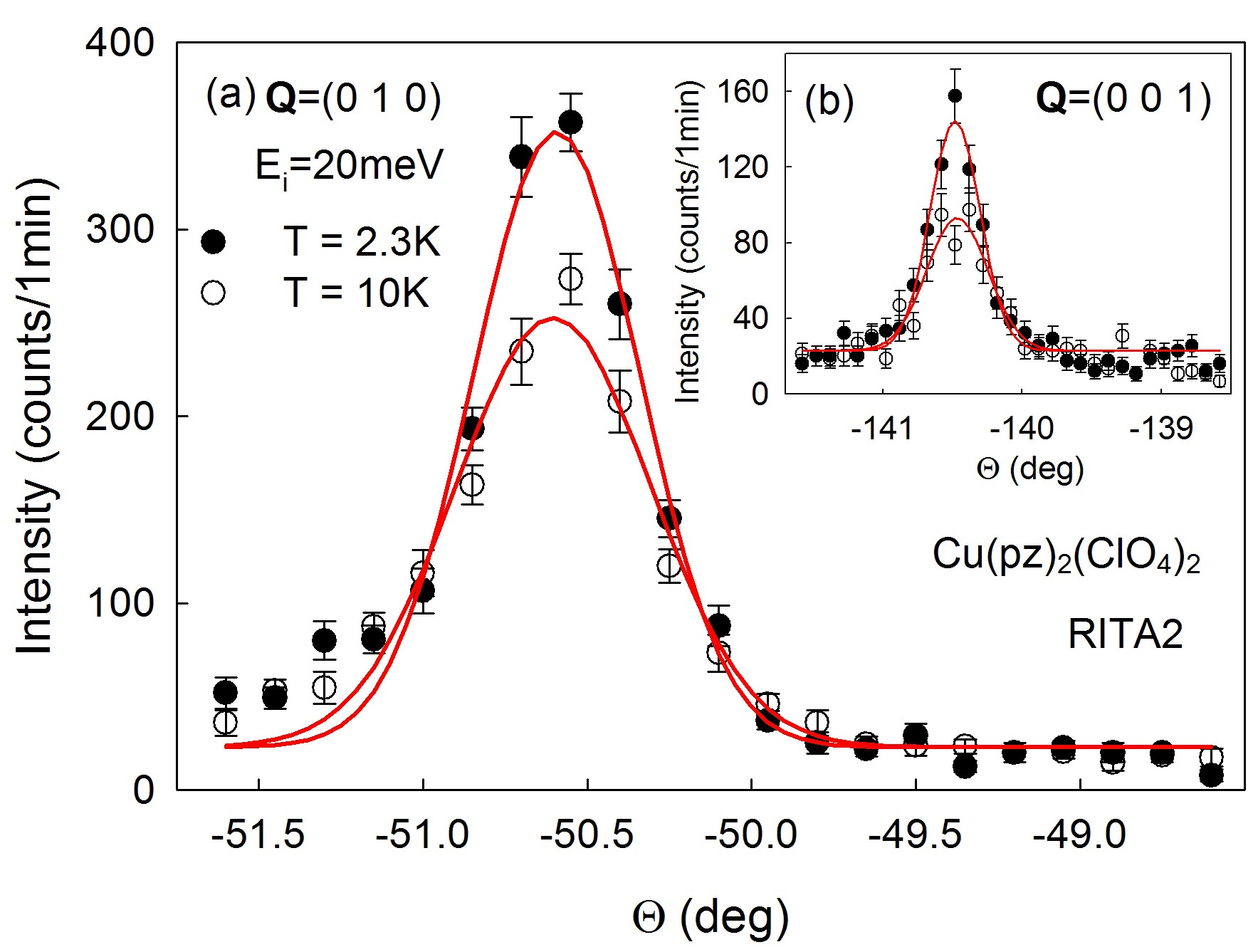}
  \caption{(Color online) (a) Neutron scattering intensity at $\textbf{Q}=(0, 1, 0)$ measured at $T=2.3\;\mathrm{K}$ and at
  $T=10\;\mathrm{K}$
as a function of a rotation of the sample around the vertical axis
described by angle $\Theta$. The inset (b) shows the neutron
scattering observed at $\textbf{Q}=(0, 0, 1)$ at the same
temperature.}
  \label{diff}
\end{center}
\end{figure}
\section{Results.}
\subsection{Specific heat measurements.}
The temperature dependence of the specific heat of \cupzclo \ is
shown in Fig.~\ref{Cp} for different magnetic fields applied
perpendicular and parallel to the copper square-lattice plane. At
all fields, the temperature dependence of the specific heat reveals
a well defined cusp-like peak, indicating a second order phase
transition towards 3D long-range magnetic order. Previous zero-field
studies of \cupzclo \ did not show an anomaly in the specific heat
\cite{Lancaster}. Most likely, the high accuracy of our measurements
played a crucial role in detecting the zero-field anomaly in the
specific-heat curve. The small size of the ordering anomaly is a
consequence of the low dimensionality of the magnetism and an
ordered magnetic moment that, due to quantum fluctuations, is
considerably smaller than the free-ion value. The HT phase diagram
assembled from the specific heat measurements is shown in
Fig.~\ref{ht}. The measurements show that the N\'{e}el temperature
increases with increasing magnetic field, from $\rm
T_N=4.24(4)\;\mathrm{K}$ at zero field to $\rm
T_N=5.59(3)\;\mathrm{K}$ at $\rm \mu_0H=9\;\mathrm{T}$.
\par
We also observe an increase of the specific heat with increasing
magnetic field in the paramagnetic phase just above the 3D ordering
temperature. We propose that the field dependence of the specific
heat data is a consequence of field-induced anisotropy in the 2D AF.
In zero field, a pure 2D Heisenberg AF orders at zero temperature,
but quantum Monte Carlo simulations \cite{Cuccoli03} have shown that
application of an external field induces an Heisenberg-XY crossover
and leads to a finite temperature Berezinskii-Kosterlitz-Thouless
transition $T_\mathrm{BKT}$ \cite{BKT1, BKT2}. One consequence of
this crossover is the increase of $T_\mathrm{BKT}$ with external
field for up to $H<H_\mathrm{SAT}/4$ and then a gradual decrease of
the transition temperature with increased fields. While the
zero-field 3D transition $T_N$ in \cupzclo \ is driven by the
combination of 3D interaction and intrinsic XY anisotropy, the
increase of $T_N$ as a function of field may thus be driven by an
increase of the effective anisotropy and the associated increase of
$T_\mathrm{BKT}$. Similarly, we propose that the increase of the
specific heat above the 3D ordering temperature is caused by the
field-induced XY anisotropy: In the 2D antiferromagnet on a square
lattice, 2D topological spin-vortices appear above the
Berezinskii-Kosterliz-Trousers (BKT) transition as the preferable
thermodynamic configuration. In applied magnetic field the vortices
unbind above the BKT transition, leading to the increase of the
specific heat above the ordering temperature. The anisotropy
crossover thus affects the specific heat in a manner similar to the
observed behavior \cite{Cuccoli03}.
\par
Remarkably, the HT phase diagrams are identical for fields parallel
and perpendicular to the square-lattice planes. This suggests that
the dominant exchange interactions between nearest copper spins
$J_1$ in $bc$-plane are close to the isotropic limit in spin space.
This should not be confused with the strong spatial
two-dimensionality of \cupzclo.
\begin{figure}
\begin{center}
  \includegraphics*[width=8cm,bbllx=0,bblly=0,bburx=1,bbury=0.74,angle=0,clip=]{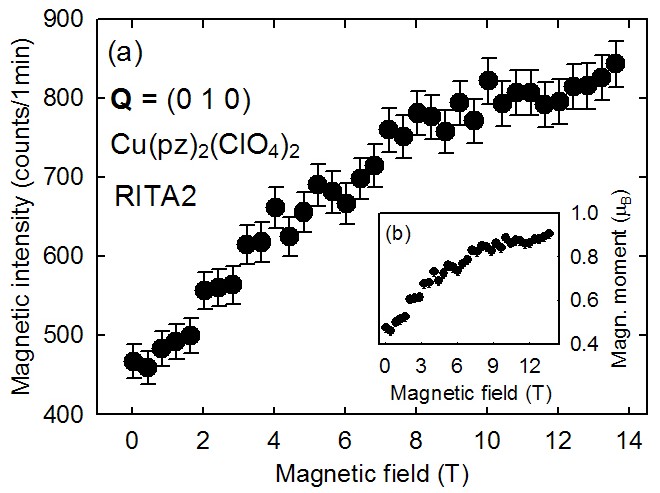}
  \caption{The magnetic peak intensity of neutron scattering at
  $\textbf{Q}=(0, 1, 0)$ as function of magnetic field measured at
$\rm T=2.3\;\mathrm{K}$. The non-magnetic scattering was estimated
from measurements at $\rm T=10\;\mathrm{K}$ and subtracted from the
overall peak intensity. The inset shows the ordered
antiferromagnetic moment
  as a function of field.}
  \label{fscan}
\end{center}
\end{figure}
\subsection{Magnetic order parameter.}
To determine the ordered magnetic structure of \cupzclo, several
magnetic Bragg reflections were measured by neutron diffraction.
Fig.~\ref{diff} shows two magnetic peaks measured above and below
the transition temperature at $\textbf{Q}=(0, 1, 0)$ and
$\textbf{Q}=(0, 0, 1)$ using a final energy $\rm
E_f=20\;\mathrm{meV}$. This data directly demonstrates the presence
of magnetic order below $\rm T_N$. The magnetic Bragg peak widths
are limited by the instrumental resolution, confirming that the
magnetic order is long-range. The field dependence of the magnetic
scattering at $\textbf{Q}=(0, 1, 0)$ measured using final energy
$\rm E_f=5\;\mathrm{meV}$ reveals an increase of magnetic intensity
as a function of field from zero to $\rm \mu_0H=13.5\;\mathrm{T}$ as
is shown in Fig.~\ref{fscan}. The magnetic scattering was determined
by subtracting the non-magnetic background determined at T = 10K.
The increase of magnetic diffraction intensity with field is most
probably related to a quenching of quantum fluctuations by the
magnetic field, that simultaneously also leads to the observed
increase of the transition temperature $\rm T_N$. This result is in
a good agreement with the specific heat data indicating enhanced XY
anisotropy in the applied magnetic field. The intensity measured at
$\textbf{Q}=(0, 1, 0)$ at $\rm T=10\;\mathrm{K}$ as the function of
applied field did not reveal any magnetic scattering, showing that
magnetic fields do not lead to field-induced antiferromagnetic order
in the paramagnetic phase.
\par
\begin{figure}
\begin{center}
  \includegraphics*[width=8.5cm,bbllx=0,bblly=0,bburx=1,bbury=0.8]{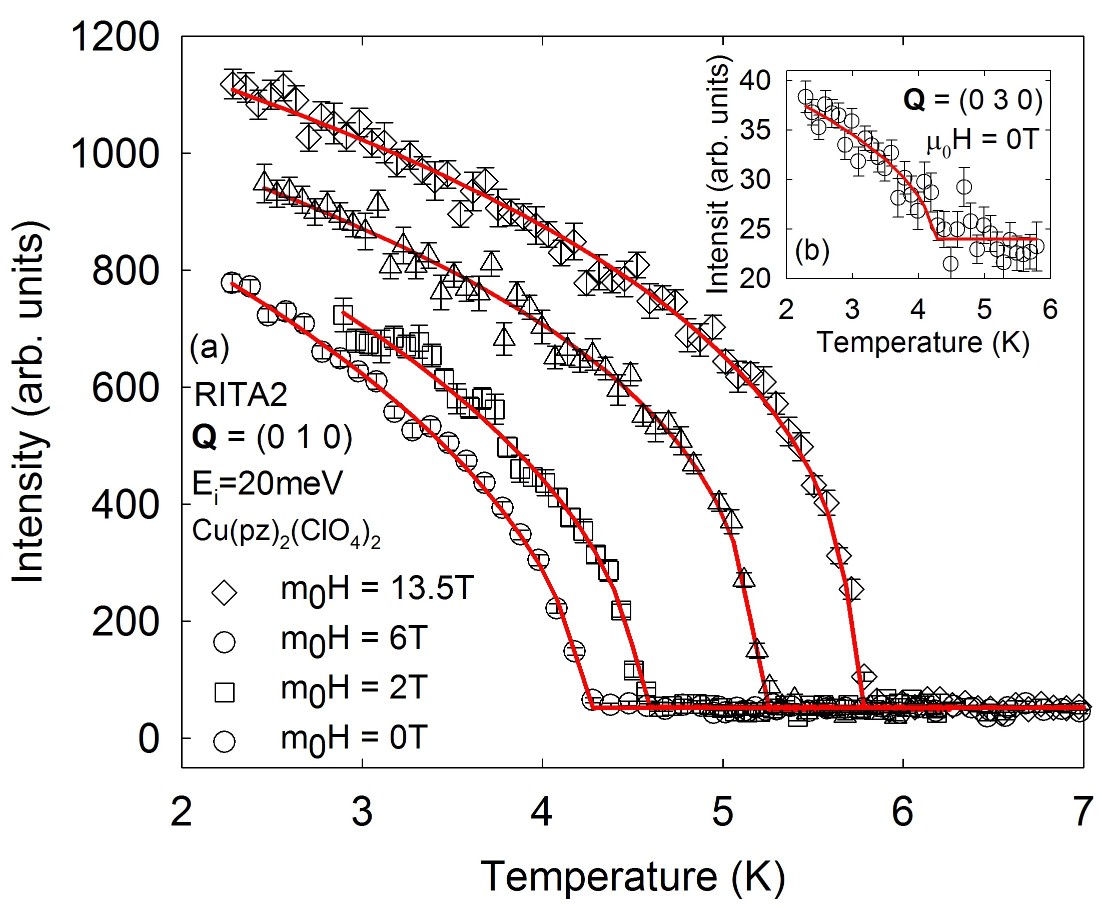}
  \caption{(Color online) (a) The temperature dependence of the neutron scattering peak intensity measured at the
antiferromagnetic point $\textbf{Q}=$(0, 1, 0). The data collected
at $\rm \mu_0H=0\;\mathrm{T}$, $\rm \mu_0H=2\;\mathrm{T}$ $\rm
\mu_0H=6\;\mathrm{T}$ and $\rm \mu_0H=13.5\;\mathrm{T}$ are shown by
circles, squares, triangles and diamonds, respectively. The red
lines are guides to the eye. The inset (b) represents the peak
intensity at $\textbf{Q}=(0,3,0)$ as the function of temperature
obtained at zero field. $\rm T_N$ was found to be the same as for
$\textbf{Q}=(0,1,0)$.}
  \label{tscans}
\end{center}
\end{figure}

The critical magnetic behavior was studied by measuring the peak
intensity of the neutron scattering at the antiferromagnetic wave
vector $\textbf{Q}=(0, 1, 0)$ and $\textbf{Q}=(0, 3, 0)$ as function
of temperature in magnetic field up to $\rm
\mu_0H=13.5\;\mathrm{T}$. Typical scans are shown in
Fig.~\ref{tscans}. The solid line display that the increase of the
antiferromagnetic intensity in the ordered phase close to $\rm T_N$
is evidently steeper for high fields. The HT phase diagram compiled
from the temperature scans is shown in Fig.~\ref{ht} and it confirms
the phase diagram obtained from specific heat measurements.
\subsection{Ordered magnetic structure.}
The symmetry of the ordered magnetic phase was studied by neutron
diffraction. Group theory was used to restrict the search only to
magnetic structures that are allowed by symmetry. The magnetic Bragg
peaks at $\textbf{Q}=(0, 1, 0)$ and $\textbf{Q}=(0, 3, 0)$ indicate
that the magnetic structure breaks the C-centering of the chemical
lattice and that \cupzclo \ adopts an antiferromagnetic structure
for $\rm T<T_N$. Symmetry analysis revealed six basis vectors which
belong to four irreducible representations and are listed in
Tab.~\ref{bv} (for details see Appendix A).
\par
The analysis is complicated by the fact that the single-crystal
probably consists of two domains with interchanged $b$- and
$c$-axis, which are nearly identical in length. A twinning of the
single-crystal in this manner is indicated by the observation of
both the $\rm \textbf{Q}=(0, 2, 3)$ and $\textbf{Q}=(0, 3, 2)$
nuclear Bragg peaks with similar intensity, although $\textbf{Q}=(0,
3, 2)$ is not allowed for a C-centered lattice.
\par
\begin{figure}
\begin{center}
  \includegraphics[width=8.5cm,bbllx=0,bblly=0,bburx=1,bbury=0.55,angle=0,clip=]{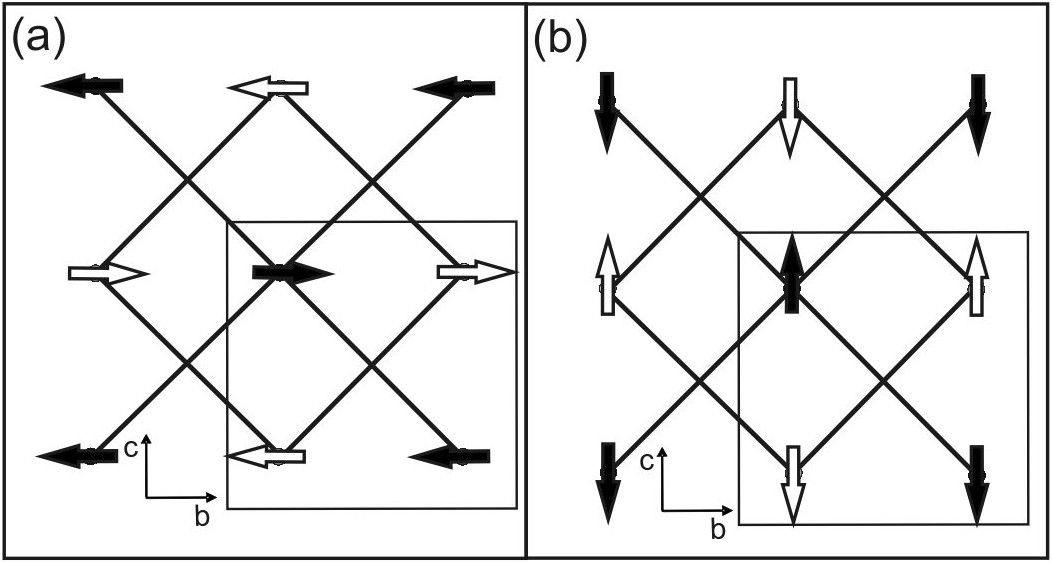}
  \caption{Two possible magnetic structures of \cupzclo \ belonging to the second
and the fourth irreducible representations (see Tab.~\ref{ct}), are
shown in (a) and (b), respectively. Two adjacent square-lattice $\rm
Cu^{2+}$ layers, separated by a (0.5,0.5,0) lattice unit
translation, are depicted by open and filled arrows. $\rm
Cu^{2+}-Cu^{2+}$ interlayer interaction pathway along $ac$ diagonal
corresponds to the vertical distance between filled and open symbols
in (a) and (b).}
  \label{spinstr}
\end{center}
\end{figure}

The experimental data are consistent with both $\Gamma_2$ and
$\Gamma_4$ irreducible representations listed in Tab.~\ref{ct} and
with two basis vectors $\overrightarrow{\phi}_2$ and
$\overrightarrow{\phi}_6$. It is not possible to distinguish between these
two solutions because $\overrightarrow{\phi}_2$ of the $bc$ crystallographic domain is identical
to $\overrightarrow{\phi}_6$ of the $cb$ crystallographic domain, and the fits were made assuming
a equal population of $bc$ and $cb$ crystallographic domains. The ordered magnetic structure of
\cupzclo \ can have magnetic moments aligned antiferromagnetically
either along crystallographic $b$- or $c$-axis as is shown in
Fig.~\ref{spinstr}(a) and Fig.~\ref{spinstr}(b), respectively. Due
to a small number of observed magnetic reflections and the
crystallographic twinning, our experiment cannot distinguish between
these two magnetic structures. The collinear spin arrangement in
$bc$-plane is consistent with the absence of the
Dzyaloshinsky-Moriya interactions between NN. The spatial
arrangement of the ordered magnetic moments in adjacent
square-lattice layers is ferro- and antiferromagnetic along $ab$-
and $ac$-diagonal, respectively. This is consistent with the
chemical structure of \cupzclo, where the interlayer interaction
pathway along $ac$-diagonal is shorter than the path along $ab$.
\par
\par
The value of the ordered magnetic moment was obtained from a
minimization of $\rm \delta=|R_{calc} - R_{exp}|$, where $\rm
R_{exp}$ is the measured ratio of the magnetic Bragg peak intensity
to the nuclear Bragg peak intensity, $\rm R_{calc}=\rm
|F(\textbf{Q})_{magn}|^2/|F(\textbf{Q})_{nucl}|^2$, $\rm
F(\textbf{Q})_{magn}$ and $\rm F(\textbf{Q})_{nucl}$ are the
magnetic and nuclear structure factors, respectively. The fit was
performed for two magnetic peaks observed at $\textbf{Q}=(0, 1, 0)$
and $\textbf{Q}=(0, 3, 0)$ and two nuclear peaks measured at
$\textbf{Q}=(0, 2, 4)$ and $\textbf{Q}=(0, 0, 6)$. The obtained value
of the ordered magnetic moment in zero field is $\rm
m_0=0.47(5)\;\mathrm{\mu_B}$. The comparison of $\rm R_{calc}$ and $\rm R_{exp}$
for
\begin{table}[ht]
\centering
 \begin{tabular}{c c c c c }
\hline \hline
& $\frac{|F(0, 1, 0)|^2}{|F(0, 2, 4)|^2}$ &   $\frac{|F(0, 1, 0)|^2}{|F(0, 0, 6)|^2}$ &  $\frac{|F(0, 3, 0)|^2}{|F(0, 2, 4)|^2}$   &  $\frac{|F(0, 3, 0)|^2}{|F(0, 0, 6)|^2}$     \\ [0.2cm]
\hline
$\rm R_{exp}\times 10^{-4}$ &  3.72(7)  &  5.13(12)  &  1.89(18)  &   2.60(25)  \\
\hline
$\rm R_{calc}\times 10^{-4}$ &  4.35  & 4.57   &  2.16 &  2.26  \\
\hline \hline
\end{tabular}
\caption{The measured and the calculated ratios of squared magnetic to nuclear structure factors for different Bragg peaks. The calculated values were obtained from a
minimization of $\rm \delta=|R_{calc} - R_{exp}|$ and correspond to the ordered magnetic moment $\rm m_0=0.47\mu_B$.}\label{Rexpcalc}
\end{table}
two magnetic and two nuclear Bragg peaks is presented in the Tab.~\ref{Rexpcalc}.
The calculated value of the ordered magnetic moment is smaller than the free-ion
magnetic moment, indicating the presence of strong quantum
fluctuations in the magnetic ground state of \cupzclo. The inset (b)
in Fig.~\ref{fscan} displays the increase of the ordered
antiferromagnetic moment from $\rm m_0=0.47(5)\mu_B$ in zero field
to $\rm m_0=0.93(5)\;\mathrm{\mu_B}$ in $\rm
\mu_0H=13.5\;\mathrm{T}$. This is direct evidence for the
suppression of quantum fluctuations by the applied magnetic field
due to induced XY anisotropy as suggested by our specific heat
measurements.
\begin{figure}
\begin{center}
  \includegraphics*[width=8.5cm,bbllx=0,bblly=0,bburx=1,bbury=1.2,angle=0,clip=]{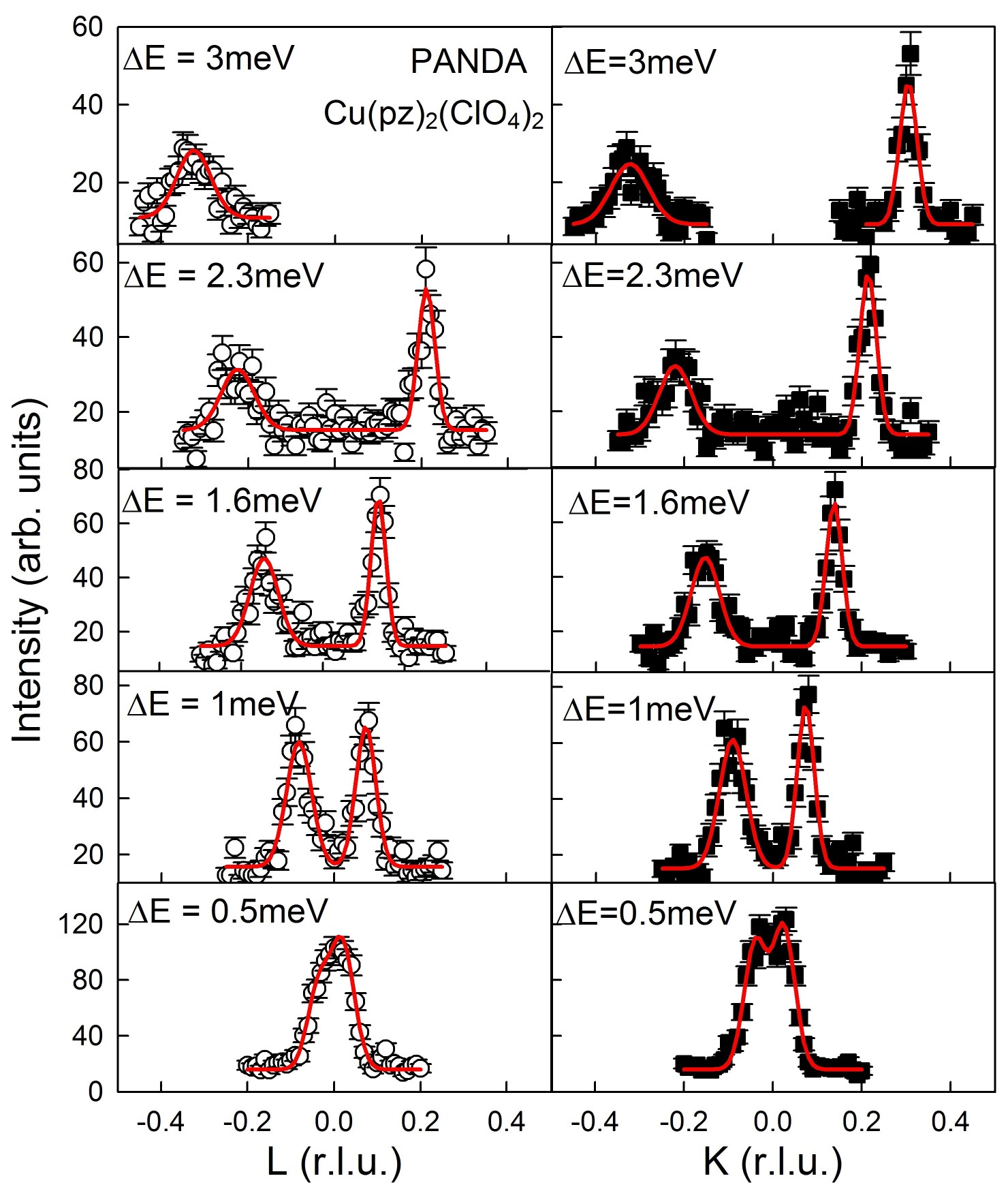}
  \caption{(Color online) A series of constant energy scans performed along the [0, k, 1]
  and [0, -1, l] directions at different energy
  transfers $\rm \Delta E$ in zero magnetic field and at $\rm T=1.42\;\mathrm{K}$. Please note the changing
scale of the vertical axis for the different scans.
  The solid lines correspond to a convolution of two Gaussians with the
resolution function.}
  \label{Qscans}
\end{center}
\end{figure}
\subsection{Spin dynamics.}
The wave-vector dependence of the magnetic excitations has been
measured using neutron spectroscopy. Constant energy scans were
performed near the antiferromagnetic zone centers $\textbf{Q}=(0, 0,
1)$ and $\textbf{Q}=(0, -1, 0)$ for energy transfer $\rm \Delta E$
in the range from $\rm \Delta E=0.5\;\mathrm{meV}$ to $\rm \Delta
E=3\;\mathrm{meV}$ and are shown in Fig.~\ref{Qscans}. The observed
magnetic peaks are resolution limited, indicating that these
magnetic excitations are long-lived magnons associated with a
long-range ordered magnetic structure.
\par
\begin{figure}[h]
\begin{center}
  \includegraphics*[width=8.5cm,bbllx=0,bblly=0,bburx=1,bbury=1,angle=0,clip=]{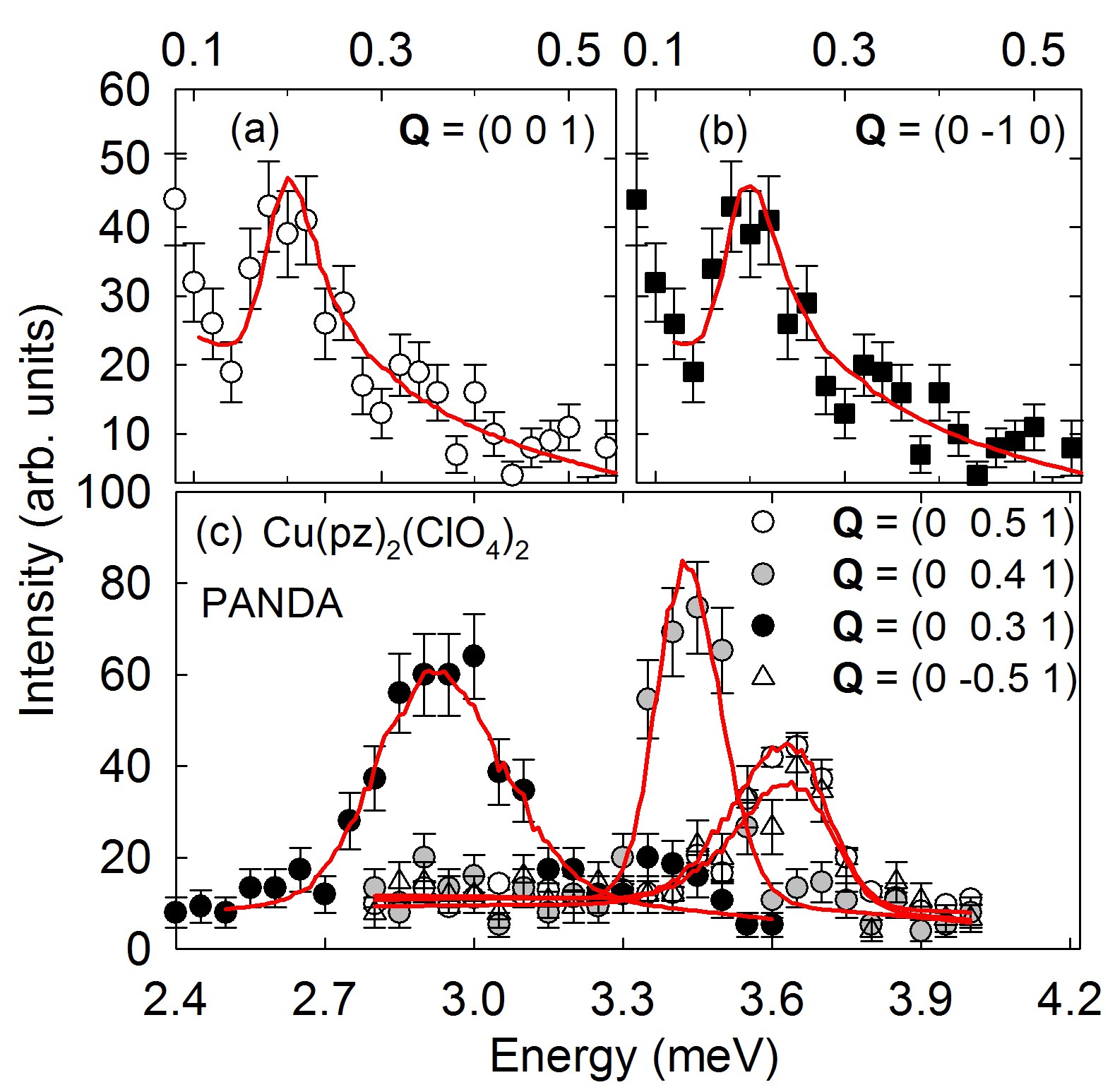}
 \caption{(Color online) The constant $\textbf{Q}$-scans collected at small energy
 transfer show the energy of a gapped spin-wave at the antiferromagnetic zone center
 performed at $\textbf{Q}=(0,0,1)$ and at $\textbf{Q}=(0,-1,0)$ are presented in (a) and (b), respectively.
 (c) Constant $\textbf{Q}$-scans performed close to the antiferromagnetic zone boundary at high energy transfer show the dispersion of the spin-wave.
The measurements were performed in zero magnetic field and at $\rm
T=1.42\;\mathrm{K}$. The solid lines represent the convolution of a
Gaussian with the resolution function.}
  \label{escans}
\end{center}
\end{figure}

Constant wave-vector scans were performed at the antiferromagnetic
zone centers in the energy transfer range from $\rm \Delta
E=0\;\mathrm{meV}$ to $\rm \Delta E=0.7\;\mathrm{meV}$
(Fig.~\ref{escans}a, b). These scans reveal a magnetic mode which is
gapped and has a finite energy $\rm E_{zc}=0.201(8)\;\mathrm{meV}$
at the antiferromagnetic zone center. The energy gap at the
antiferromagnetic zone center is attributed to the presence of a
small XY anisotropy in the nearest-neighbor two-ion exchange
interactions, because a single-ion anisotropy of type $\rm
\textbf{D}(\textbf{S}^z)^2$ is not allowed for S=1/2.
\par
Constant wave-vector scans away from the antiferromagnetic zone
center carried out at higher energies are shown in
Fig.~\ref{escans}(c). The energies of the magnetic excitation at the
symmetrically identical antiferromagnetic zone boundary points $\rm
\textbf{Q}_{zb1}=(0, 0.5, 1)$ and $\rm \textbf{Q}_{zb2}=(0, -0.5,
1)$ are equal to $\rm E_{zb1}=3.629(6)\;\mathrm{meV}$ and $\rm
E_{zb2}=3.599(13)\;\mathrm{meV}$, respectively. The peaks observed
in the constant wave-vector scans at $\rm \textbf{Q}_{zb1}$ and $\rm
\textbf{Q}_{zb2}$ are resolution limited. This experimental fact
together with the identity of the values $\rm E_{zb1}$ and $\rm
E_{zb2}$ confirms the NN interactions in $bc$-plane are identical
along the square-lattice directions. In case of different strengths
for the NN interactions in $bc$-plane a broadening of the peaks at
$\rm \textbf{Q}_{zb1}$ and $\rm \textbf{Q}_{zb2}$ would be observed.
From our previous study \cite{tsyrulin} we know that there is also a
small NNN interaction equal to $2\%$ of the NN exchange interaction.
Therefore the observed one-magnon mode was compared to the following
model 2D Hamiltonian:
\begin{equation}
     \rm \hat{H}=\sum_{\langle i,j\rangle}\{J_1^z\textbf{S}_i^z\cdot\textbf{S}_j^z
     +J_1^{xy}(\textbf{S}_i^x\cdot\textbf{S}_j^x+\textbf{S}_i^y\cdot\textbf{S}_j^y)\} +
     J_2 \sum_{\langle i,k \rangle}\textbf{S}_i\cdot\textbf{S}_k,
\label{Hnnn}
\end{equation}
where $\rm \langle i,j\rangle$ indicates the sum over NN in the
$bc$-plane, $\rm \langle i,k\rangle$ - the sum over NNN in the
$bc$-plane, $\rm J_1^z$, $\rm J_1^{xy}$ and $\rm J_2$ are $z$-,
$xy$-components of the NN interaction and the NNN exchange,
respectively.
\par
The linear spin wave theory (for details see Appendix B) yields two
spin-wave modes with the dispersion $\rm
\hbar\omega_q=\sqrt{A_q^{2}-B_q^{2}}$, where $\rm
A_q=4SJ_1^{xy}+S(J_1^{xy}-J_1^z)(\cos(\textbf{qb})+\cos(\textbf{qc}))
-4SJ_2+4SJ_2\cos(\textbf{qb})\cdot\cos(\textbf{qc}))$ and $\rm
B_q=S(J_1^{xy}+J_1^z)(\cos(\textbf{qb})+\cos(\textbf{qc}))$. This
implies that the exchange anisotropy mostly affects the magnon
energy close to the antiferromagnetic zone center, while the zone
boundary energy remains nearly unaffected by the exchange
anisotropy. In the 2D S=1/2 AF, the energy of a classical (large-S)
spin-wave mode is renormalized due to quantum fluctuations with the
best theoretically predicted renormalization factor $\rm Z_c=1.18$
\cite{igarashi, singh}. Therefore the energy at the
antiferromagnetic zone boundary is equal to $\rm
E_{zb}=2Z_cJ_1^z-J_{2R}$, where $\rm J_{2R}$ is the renormalized NNN
interaction. The calculated $xy$-component of NN and NNN exchange
interactions are equal to $\rm J_1^{xy}=1.563(13)\;\mathrm{meV}$ and
$\rm J_2 \simeq 0.02J_1^{xy}$, respectively. According to the linear
spin wave theory $\rm E_{zc}^2=8J_1^{xy}(J_1^{xy}-J_1^z)$ and thus
$\rm J_1^{z}=0.9979(2)J_1^{xy}$.
\par
The values of the $xy$- and $z$-components of the NN interaction
obtained from our neutron measurements are in a good agreement with
the result of magnetic susceptibility measurements \cite{Xiao},
which yielded $\rm J_1^{xy}=1.507(26)\;\mathrm{meV}$ and $\rm
J_1^{z}=0.9954J_1^{xy}$. The small XY anisotropy indicates that the
dominant exchange interaction between nearest copper ions in
$bc$-plane in \cupzclo, while spatially very anisotropic, is close
to the isotropic limit in spin space, explaining the strong
similarity of the HT phase diagrams measured in magnetic fields
applied parallel and perpendicular to copper square-lattice
(Fig.~\ref{ht}).
\par
\begin{figure}
\begin{center}
  \includegraphics*[width=8.5cm,bbllx=0,bblly=0,bburx=1,
  bbury=0.8,angle=0]{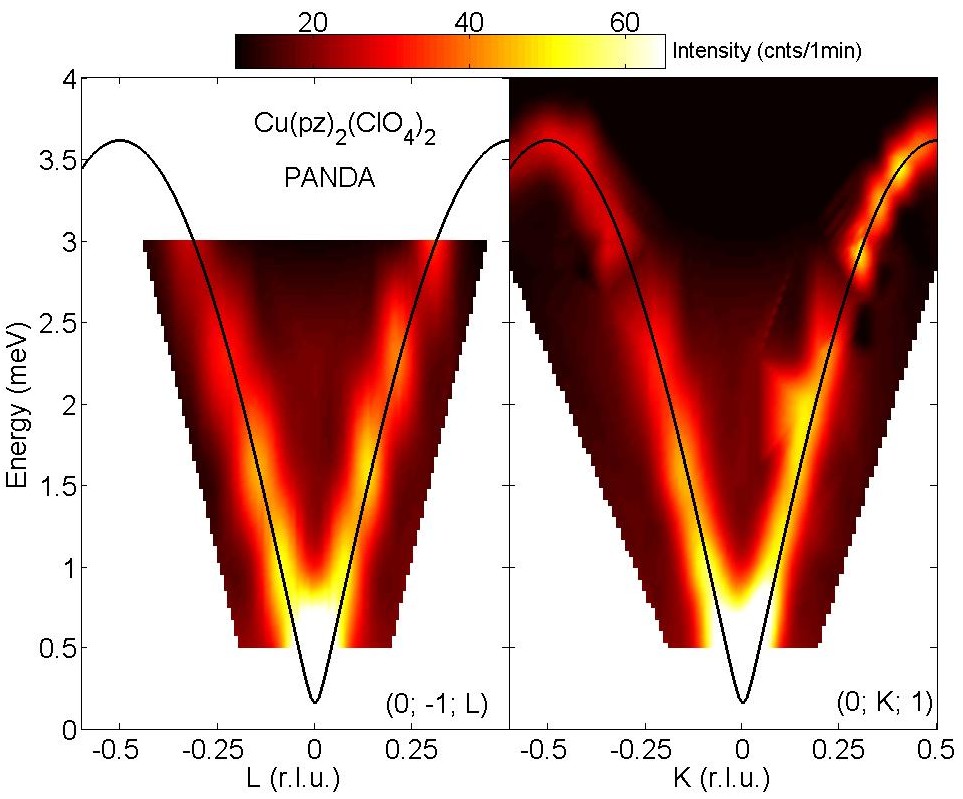}
  \caption{(Color online) Color plot of the scattering intensity, showing the dispersion
  along the $\textbf{Q}=(0, -1, l)$ and $\textbf{Q}=(0, k, 1)$ directions measured at zero field, presented in the left and right panels, respectively.
  The color plot was obtained by merging a total of five and thirteen constant wave-vector scans, respectively. The solid line represents the
  dispersion computed from linear spin wave theory using
  $\rm J_1^{xy}=1.563\;\mathrm{meV}$, $\rm J_1^{z}=0.9979J_1^{xy}$ and NNN exchange equal to $\rm J_2=0.02J_1^{xy}$ as described in the text.}
  \label{cplot}
\end{center}
\end{figure}

The inelastic-scattering data were fitted with the Gaussian
instrumental resolution function convoluted numerically with the
model Hamiltonian (\ref{Hnnn}). The result of the fits is shown by
the red lines in Fig.~\ref{Qscans} and Fig.~\ref{escans}, and
provides a good description of the observed spin waves. The color
plot of the neutron scattering intensity, which is shown in
Fig.~\ref{cplot}, summarizes the observed magnetic excitations in
both crystallographic directions. The black lines display the result
of the linear spin wave theory, showing that the observed dispersive
excitation is well characterized by the Hamiltonian (\ref{Hnnn}).
\par
The measured spin wave dispersion is similar to that observed in
another 2D square-lattice antiferromagnetic material, namely copper
deuteroformate tetradeuterate (CFTD), where the exchange interaction
strength is equal to $\rm J=6.3(3)\;\mathrm{meV}$ and the energy gap
of $\rm E=0.38(2)meV$ is present at the antiferromagnetic zone
center. However, the energy gap in CFTD is induced by the presence
of small antisymmetric Dzyaloshinsky-Moriya interaction $\rm
\textbf{D}=0.0051(5)\;\mathrm{meV}$ between NN \cite{clarke2,
ronnow}. Another example with comparable properties is
$\rm{K_2V_3O_8}$ with 2D NN exchange strength $\rm
J=1.08(3)\;\mathrm{meV}$ and small energy gap at antiferromagnetic
point equal to $\rm E=0.072(9)\;\mathrm{meV}$ \cite{lumsden}, which
is described by Dzyaloshinsky-Moriya and easy-axis anisotropies
\cite{Chernyshev}. In contrast, Dzyaloshinsky-Moriya interactions
between NN in \cupzclo \ are forbidden by symmetry and the energy
gap at the antiferromagnetic zone center is generated by small XY
anisotropy.
\par
\begin{figure}
\begin{center}
  \includegraphics*[width=8.5cm,bbllx=0,bblly=0,bburx=1,
  bbury=0.7,angle=0]{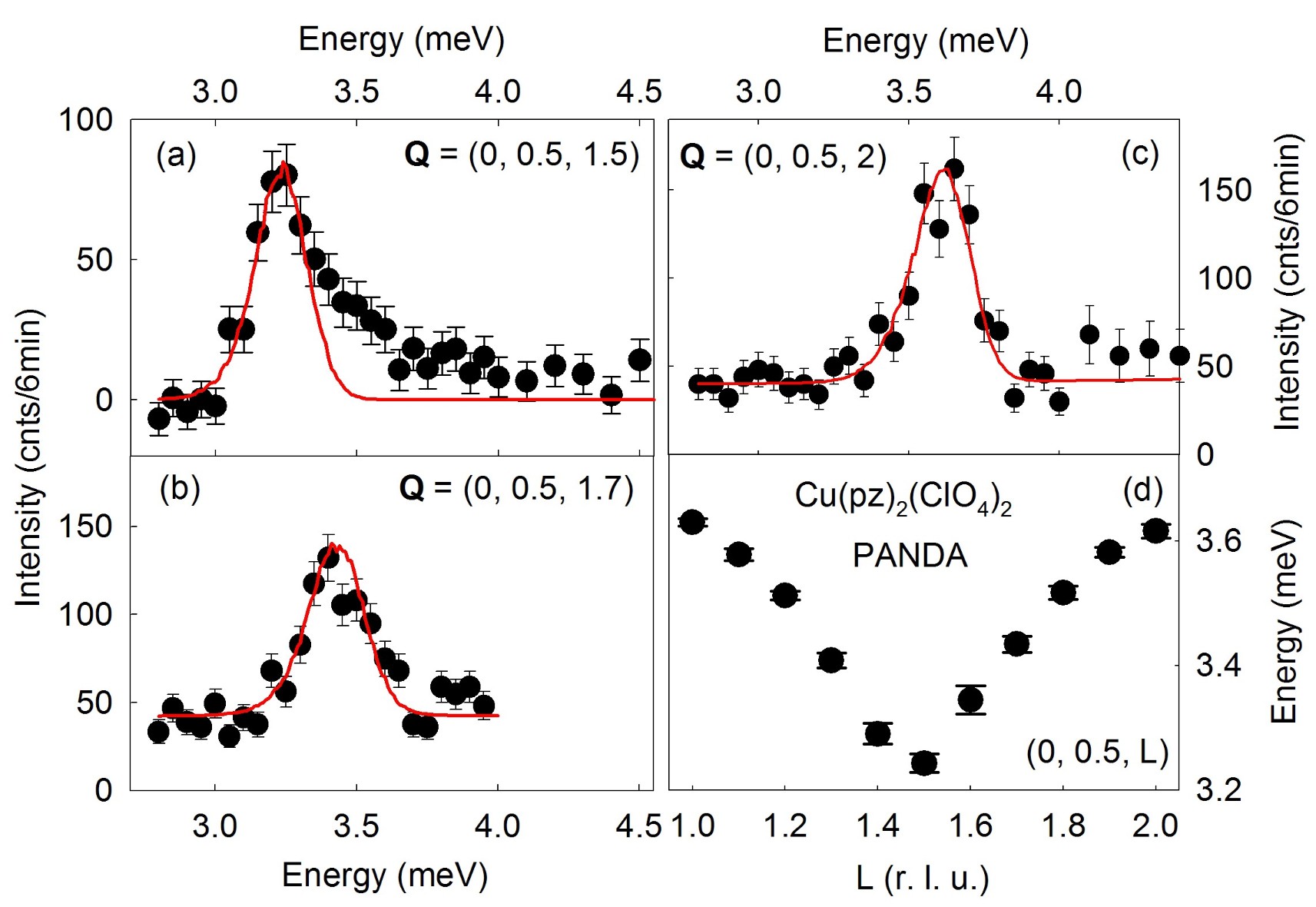}
  \caption{(Color online) (a) The energy scan at $\textbf{Q}=(0, 0.5,
1.5)$ with the background subtracted as explained in the text. The
energy scans performed at wave-vectors $\textbf{Q}=(0, 0.5, 1.7)$
and $\textbf{Q}=(0, 0.5, 2)$ are shown in (b) and (c), respectively.
The red curves are fits of a Gaussian function convoluted with the
resolution function. (d) The antiferromagnetic zone boundary
dispersion measured in zero magnetic field and at $\rm
T=1.42\;\mathrm{K}$.}
  \label{zb}
\end{center}
\end{figure}

We also studied the spin-wave dynamics along the antiferromagnetic
zone boundary by performing constant wave-vector scans along the
$\textbf{Q}=(0, 0.5, l)$ direction from $l=1$ to $l=2$. Typical data
are shown in Fig.~\ref{zb}(a-c) and the observed zone boundary
dispersion is shown in Fig.~\ref{zb}(d). The onset of the scattering
at $\textbf{Q}=(0, 0.5, 1.5)$ is reduced by 10.7(4)\% in energy
compared to $\textbf{Q}=(0, 0.5, 1)$, confirming our recent
independent measurement \cite{tsyrulin}. The decrease of the
resonant mode energy at $\textbf{Q}=(0, 0.5, 1.5)$ results from a
resonating valence bond quantum fluctuations between NN spins
\cite{Sandvik,McMorrow}. The observed dispersion at the zone
boundary is slightly larger than expected from series expansion
calculations and Quantum Monte Carlo simulations for 2D Heisenberg
square-lattice AF with NN interactions \cite{Sandvik, Zheng} and can
be explained by the presence of a small antiferromagnetic NNN
interaction.
\par
In order to subtract a nonmagnetic contribution from the background
in the energy scan at $\textbf{Q}=(0, 0.5, 1.5)$ we performed
measurements with the sample turned away from magnetic scattering.
The background-subtracted data are shown in Fig.~\ref{zb}(a). The
width of the scattering peak as a function of energy at
$\textbf{Q}=(0, 0.5, 1.5)$ is clearly broader than the instrumental
resolution. This implies the existence of a magnetic continuum
scattering in this region of the antiferromagnetic zone boundary.
The magnetic continuum with the present PANDA measurements is
consistent with our previous investigation of \cupzclo \
\cite{tsyrulin}. This non-trivial magnetic continuum and the
dispersion at the zone-boundary result from quantum fluctuations in
\cupzclo \ which are enhanced by a small antiferromagnetic NNN
interaction.
\section{Conclusions.}
We have performed a comprehensive study of the 2D S=1/2
square-lattice AF \cupzclo \ using a collection of experimental
techniques, including specific heat measurements, neutron
diffraction and neutron spectroscopy. The HT phase diagram was
mapped out for magnetic fields up to $\rm \mu_0H=9T$ applied
parallel and perpendicular to the $\rm Cu^{2+}$ square-lattice
planes, showing that the HT boundaries of the ordered phase have the
same field dependence. This result shows that the dominant exchange
interactions between nearest spins are close to the isotropic limit.
Applied magnetic fields induces XY spin anisotropy leading to quench
of quantum fluctuations, as we observed in the specific heat
measurements.
\par
Neutron diffraction confirms the HT phase diagram obtained from
macroscopic measurements and extends it up to $\rm
\mu_0H=13.5\;\mathrm{T}$. The ordered magnetic moment of $\rm
Cu^{2+}$ ions at zero field, $\rm m_0=0.47(5)\;\mathrm{\mu_B}$, is
reduced from the expected value, indicating the existence of strong
quantum fluctuations in the ground state. Magnetic fields quench
quantum fluctuations and significantly increase the size of the
ordered magnetic moment to $\rm m_0=0.93(5)\;\mathrm{\mu_B}$ at $\rm
\mu_0H=13.5\;\mathrm{T}$. Therefore neutron diffraction confirms the
suppression of quantum fluctuations by field-induced anisotropy seen
in our specific heat measurements. The study of the ordered magnetic
structure by neutron diffraction shows 2D antiparallel alignment of
spins at the neighbor sites.
\par
At zero field, we observed a well-defined magnon mode using neutron
spectroscopy. Its dispersion is described by the spin Hamiltonian
with slightly anisotropic NN interaction with $xy$- and
$z$-components equal to $\rm J_1^{xy}=1.563(13)\;\mathrm{meV}$, $\rm
J_1^{z}=0.9979(2)J_1^{xy}$ and NNN exchange equal to $\rm J_2\simeq
0.02J_1^{xy}$. Therefore the closeness of the dominant exchange
interaction $\rm J_1$ to the isotropic limit measured by neutron
spectroscopy explains the similarity of the HT phase diagrams
obtained by specific heat measurements in magnetic fields applied
parallel and perpendicular to $bc$-plane.
\par
Our results obtained by three different experimental techniques
confirm and supplement each other clearly demonstrating that
\cupzclo \ is the first weakly frustrated 2D S=1/2 AF on a square
lattice with the absence of Dzyaloshinsky-Moriya interaction between
NN. The measurements verify a relatively large 10.7(4)\%
zone-boundary dispersion and a rather strong magnetic continuum at
the zone boundary. We associate these features with a resonating
valence bond fluctuations which are enhanced by a small NNN AF
interactions.
\section{Acknowledgments.}
It is a pleasure to thank Dirk Etzdorf and Harald Schneider for
technical assistance. This work was supported by the Swiss NSF
(Contract No. PP002-102831).
\section{Appendix A: The group theory analysis.}
\begin{table}[ht]
\centering
 \begin{tabular}{|c|c|c|c|c|}
\hline
 & $1$ & $2_b$& $\overline{1}$ & $m_{ac}$ \\
\hline
$\Gamma_1$ &  1  &  1  &  1  &  1 \\
\hline
$\Gamma_2$ &  1  &  1  & -1  & -1 \\
\hline
$\Gamma_3$ &  1  & -1  &  1  & -1 \\
\hline
$\Gamma_4$ &  1  & -1  & -1  &  1 \\
\hline
\end{tabular}
\caption{The character table and the irreducible representations
obtained by performing group theory analysis for monoclinic space
group C2/c ($\#15$), the table setting choice is b1) and the
magnetic ordering vectors $\textbf{k}=(0, 0, 1)$ and $\textbf{k}=(0, 0, 0)$.} \label{ct}
\end{table}
The crystal structure of \cupzclo \ belongs to the monoclinic C2/c
space group ($\#15$), whose Laue class and the point group are 2/m.
The ${\rm Cu^{2+}}$ ions occupy 4e Wyckoff positions and they are
located at $\rm r_1=(0 \ 0.7499 \ 0.25)$, $\rm r_2=(0 \ 0.2501 \
0.75)$, $\rm r_3=(0.5 \ 0.2499 \ 0.25)$  and  $\rm r_4=(0.5 \ 0.7501
\ 0.75)$. For C2/c space group, reciprocal lattice points are
located at ${\bf Q}=(h,k,l)$ with $h+k=2n$. The magnetic Bragg peaks
were observed at $\textbf{Q}_1=(0, 1, 0)$ and at $\textbf{Q}_2=(0, 0, 1)$
indicating that only $\textbf{Q}_2$ can be the magnetic ordering vector.
However,
taking into account nearly identical crystallographic parameters $b$ and $c$ and the presence
of $bc$ and $cb$ crystallographic domains, the magnetic Bragg peaks at $\textbf{Q}_1$ and $\textbf{Q}_2$
are indistinguishable. Therefore we can not identify whether the magnetic ordering vector is $\textbf{k}=(0, 0, 1)$ or $\textbf{k}=(0, 0, 0)$.
The subgroups of the ordering wave vectors are identical and consist of
four symmetry operations which belong to four different classes:
$1$, $2_b$,$\overline{1}$ and $m_{ac}$. Here, $1$ is the identity,
$2_b$ is a two-fold rotation around the b-axis, $\overline{1}$ is
the inversion and $m_{ac}$ is a mirror plane in the ac plane.
Therefore, there are four one-dimensional irreducible
representations whose characters are summarized in the character
table given in Tab.~\ref{ct}. The decomposition equation for the
magnetic representation is $\rm
\Gamma_{mag}=1\Gamma_1+1\Gamma_2+2\Gamma_3+2\Gamma_4$. The six basis
vectors presented in Tab.~\ref{bv} are calculated for two $\rm
Cu^{2+}$ positions in primitive unit cell using the projection
operator method acting on a trial vector $\phi_{\alpha}$
$$
\rm \Psi_{\alpha\nu}^{\lambda} = \sum_{g\epsilon G_k}
D_{\nu}^{\lambda *}(g) \sum_{i} \delta_{i,g_i} R_g \phi_{\alpha}
det(R_g),
$$
where $\rm \Psi_{\alpha\nu}^{\lambda}$ is the basis vector projected
from the $\rm \lambda^{th}$ row of the $\rm \nu^{th}$ irreducible
representation, $\rm D_{\nu}^{\lambda *}(g)$ is the $\rm
\lambda^{th}$ row of the matrix representative of the $\rm \nu^{th}$
irreducible representation for symmetry operation g, i denotes the
atomic position and $\rm R_g$ is the rotational part of the symmetry
operation g. Note that basis vectors are identical for both $\textbf{k}=(0, 0, 1)$
and $\textbf{k}=(0, 0, 0)$.
\par
Dzyaloshinsky-Moriya (DM) interactions are defined as
$$
\rm \hat{H}_{DM}=\sum_{\langle i,j\rangle}\ \textbf{D}_{ij}\cdot
[\textbf{S}_i\times\textbf{S}_j],
$$
where $\rm \textbf{D}_{ij}$ is an axial vector. Action of any
symmetry operation (including lattice translations) $A$ on a DM
vector $\rm \textbf{D}_{ij}$ must be equal to $\rm
\textbf{D}_{A(i)A(j)}$ and $\rm \textbf{D}_{ij}=-\textbf{D}_{ji}$.
We analyze the action of the inversion symmetry operation
$\overline{1}$ on the axial DM vector $\rm \textbf{D}_{12}$, where
$i=1$ and $j=2$ denotes the NN copper positions $\rm r_1=(0 \ 0.7499
\ 0.25)$ and $\rm r_2=(0 \ 0.2501 \ 0.75)$, respectively. The result
of operation is
$$
\overline{1}(\rm \textbf{D}_{12})=(D_{12}^x \ D_{12}^y \ D_{12}^z).
$$
The application of the inversion symmetry on the ions positions
leads to
$$
\overline{1}(\rm \textbf{D}_{12})=\rm
\textbf{D}_{\overline{1}(1)\overline{1}(2)}=(D_{21}^x \ D_{21}^y \
D_{21}^z).
$$
These relations imply $\rm (D_{12}^x \ D_{12}^y \
D_{12}^z)=(D_{21}^x \ D_{21}^y \ D_{21}^z)$ which is possible only
in case of $\rm \textbf{D}_{12}=0$. Therefore, DM interactions
between NN in \cupzclo \ are forbidden by the crystal symmetry.
\section{Appendix B: The linear spin wave theory.}
Assuming the system in the antiferromagnetic N\`{e}el ground state
with spins pointing along $z$ and $-z$ direction we can make the
Holstein-Primakoff transformation of the spin compounds into bosonic
creation and annihilation operators. In linear spin wave
approximation it gives:
$$
    \rm S^{z}_{i}=S-a_{i}^{+}a_{i}, \ S^{z}_{j}=-S+a_{j}^{+}a_{j}, \ S^{z}_{k}=S-a_{k}^{+}a_{k},
$$
$$
    \rm S^{x}_{i}=\sqrt{2S} \ \frac{a_{i}+a_{i}^{+}}{2}, \ S^{y}_{i}=\sqrt{2S} \ \frac{a_{i}-a_{i}^{+}}{2i},
$$
$$
    \rm  S^{x}_{j}=\sqrt{2S} \ \frac{a_{j}+a_{j}^{+}}{2}, \ S^{y}_{j}=\sqrt{2S} \ \frac{-a_{j}+a_{j}^{+}}{2i},
$$
$$
\rm S^{x}_{k}=\sqrt{2S} \ \frac{a_{k}+a_{k}^{+}}{2},\
S^{y}_{k}=\sqrt{2S} \ \frac{a_{k}-a_{k}^{+}}{2i}.
$$
\begin{table}
\begin{tabular}{|l|l|l|l|l|l|}
\hline

\multicolumn{2}{|c|}{} & $X_1$ & $X_2$\\
\hline
$\Gamma_1$  & $\overrightarrow{\phi}_1$ & (0 1 0) & (0 1 0)\\
\hline
$\Gamma_2$  & $\overrightarrow{\phi}_2$ & (0 1 0) & (0 -1 0)\\
\hline
$\Gamma_3$  & $\overrightarrow{\phi}_3$ & (1 0 0) & (1 0 0)\\
           & $\overrightarrow{\phi}_4$ & (0 0 1) & (0 0 1)\\
\hline
$\Gamma_4$  & $\overrightarrow{\phi}_5$ & (1 0 0) & (-1 0 0)\\
           & $\overrightarrow{\phi}_6$ & (0 0 1) & (0 0 -1)\\
\hline
\end{tabular}
\caption{Six basis vectors calculated for two $\rm Cu^{2+}$
positions in primitive unit cell as explained in Appendix A.}
\label{bv}
\end{table}
The quantization axis lies in $bc$-plane and therefore $\rm
J_1^x=J_1^z=J, J_1^y=J-\Delta$. The spin Hamiltonian written in
bosonic operators is:

$$
     \rm \hat{H}=\sum_{\langle i,j\rangle}\{4JS(a_{i}^{+}a_{i}+a_{j}^{+}a_{j})
 + 2S\frac{2J+\Delta}{2}(a_{i}a_{j}+a_{i}^{+}a_{j}^{+}) +
$$
$$
 \rm 2S\frac{\Delta}{2}(a_{i}a_{j}^{+}+a_{i}^{+}a_{j})\} +
$$
$$
     \rm \sum_{\langle
     i,k\rangle}4J_2S\{(a_{i}a_{k}^{+}+a_{i}^{+}a_{k})-(a_{i}^{+}a_{i}+a_{k}^{+}a_{k})\},
$$
where $\rm \langle i,j\rangle$ indicates the sum over NN in the
bc-plane, $\rm \langle i,k\rangle$ - the sum over NNN in the
bc-plane. After Fourier transformation obtained Hamiltonian can be
diagonalized using standard Bogoliubov transformation:
$$
\rm a_{q}=-u_{q}\alpha_{q}+v_{q}\beta_{q}^{+},
$$
$$
\rm a_{q}^{+}=-u_{q}\alpha_{q}^{+}+v_{q}\beta_{q},
$$
$$
\rm a_{-q}=v_{q}\alpha_{q}^{+}-u_{q}\beta_{q},
$$
$$
\rm a_{-q}^{+}=v_{q}\alpha_{q}-u_{q}\beta_{q}^{+},
$$
where $\rm \alpha_{q}$ and $\rm \beta_{q}$ are the bosonic operators
and $\rm v_q$, $\rm u_q$ are numbers. Finally, after the
diagonalization we have
\begin{equation}
     \rm
\hat{H}_q=E_{g.s.}+\sum_q
S[2A_q\alpha_{q}^{+}\alpha_{q}+B_q(\alpha_{q}^{+}\alpha_{-q}^{+}+\alpha_{q}\alpha_{-q})],
\label{Hf}
\end{equation}
and the eigenstates are given by $\rm
\hbar\omega_q=(A_q^2-B_q^2)^{1/2}$, where
$$
\rm A_q=4SJ+S\Delta(\cos(\textbf{qb})+\cos(\textbf{qc}))-
$$
$$
\rm 4SJ_2+4SJ_2\cdot\cos(\textbf{qb})\cdot\cos(\textbf{qc})
$$
and
$$
\rm B_q=2S(J-1/2\Delta)(\cos(\textbf{qb})+\cos(\textbf{qc})).
$$


\begin{thebibliography}{10}

\bibitem{Haldane}F.~D.~M.~Haldane, Phys Lett. A {\bf93}, 464 (1983); F.~D.~M.~Haldane,
Phys. Rev. Lett. {\bf50}, 1153 (1983).

\bibitem{Buyers}W.~J.~L~Buyers, R.~M.~Morra, R.~L.~Armstrong, M.~J.~Hogan,
P.~Gerlach and K.~Hirakawa, Phys. Rev. Lett. {\bf56}, 371-374
(1986).

\bibitem{Tennant}D.~A.~Tennant, R.~A.~Cowley, S.~E.~Nagler and
A.~M.~Tsvelik Phys. Rev. B {\bf52}, 13368 (1995).

\bibitem{Singh}R.~R.~P.~Singh and M.~P.~Gelfand, Phys. Rev. B {\bf52}, R15695
(1995).

\bibitem{Sandvik}A.~W.~Sandvik, R.~R.~P.~Singh, Phys. Rev. Lett {\bf86}, 528
(2001).

\bibitem{Zheng}W.~Zheng, J.~Oitmaa, and C.~J.~Hamer, Phys. Rev. B {\bf71},
184440 (2005).

\bibitem{Ho}C.~M.~Ho, V.~N.~Muthukumar, M.~Ogata, and P.~W.~Anderson, Phys. Rev. Lett. {\bf86}, 1626
(2001).

\bibitem{igarashi}J.~I.~Igarashi, Phys. Rev. B {\bf46}, 10763 (1992).

\bibitem{singh}R.~R.~P.~Singh and M.~P.~Gelfand, Phys. Rev. B {\bf52}, R15695 (1995).

\bibitem{clarke2}S.~J.~Clarke, A.~Harrison, T.~E.~Mason, D.~Visser, Solid State Commun., {\bf112}, 561-564 (1999).

\bibitem{ronnow}H.~M.~R{\o}nnow, D.~F. McMorrow, R.~Coldea, A.~Harrison, I.~D.~Youngson, T.~G.~Perring,
G.~Aeppli, O.~Syljuasen, K.~Lefmann, and C.~Rischel, Phys. Rev.
Lett., {\bf87}, 037202 (2001).

\bibitem{Harris}A.~B.~Harris, A.~Aharony, O.~Entin-Wohlman, I.~Ya.~Korenblit, R.~J.~Birgeneau and Y.-J.~Kim,
Phys. Rev. B {\bf64} 024436 (2001).

\bibitem{Christensen}N.~B.~Christensen, D.~F.~McMorrow, H.~M.~Ronnow, A.~Harrison, T.~G.~Perring,
and R.~Coldea, J. Mag. Magn. Mater. {\bf272-276}, 896 (2004).

\bibitem{lumsden}M.~D.~Lumsden, S.~E.~Nagler, B.~C.~Sales, D.~A.~Tennant, D.~F.~McMorrow,
S.-H.~Lee and S.~Park, Phys. Rev. B {\bf74}, 214424 (2006).

\bibitem{McMorrow}N.~B.~Christensen, H.~M.~Ronnow, D.~F.~McMorrow, A.~Harrison, T.~G.~Perring,
M.~Enderle, R.~Coldea, L.~P.~Regnault and G.~Aeppli, Proc. Natl.
Acad. Sci. U.S.A., Vol. 104 (39), pp. 15264-15269 (2007).

\bibitem{Coldea}R.~Coldea, S.~M.~Hayden, G.~Aeppli, T.~G.~Perring,
C.~D.~Frost, T.~E.~Mason, S.-W.~Cheong, and Z. Fisk, Phys. Rev.
Lett. {\bf 86}, 5377 (2001).

\bibitem{Chandra}P.~Chandra and B.~Doucot, Phys. Rev. B {\bf 38}, 9335 (1988).

\bibitem{Chandra2}P.~Chandra, P.~Coleman and A.~I.~Larkin,  Phys. Rev. Lett. {\bf 64}, 88 - 91
(1990).

\bibitem{Read}N.~Read and S.~Sachdev, Phys. Rev. Lett. {\bf 66}, 1773
(1991).

\bibitem{Viana}J.~R.~Viana and J.~R.~de~Sousa, Phys. Rev. B {\bf 75}, 052403
(2007).

\bibitem{tsyrulin}N.~Tsyrulin, T.~Pardini, R.~R.~P.~Singh, F.~Xiao, P.~Link,
A.~Schneidewind, A.~Hiess, C.~P.~Landee, M.~M.~Turnbull, and
M.~Kenzelmann, Phys. Rev. Lett.,  {\bf 102}, 197201 (2009).

\bibitem{Wo}F.~M.~Woodward, P.~J.~Gibson, G.~Jameson, C.~P.~Landee, M.~M.~Turnbull
and R.~D.~Willett, Inorg. Chem., {\bf 46}, 4256-4266 (2007).

\bibitem{Lancaster}T. Lancaster, S.~J.~Blundell, M.~L.~Brooks, P.~J.~Baker,
F.~L.~Pratt, J.~L.~Manson, M.~M.~Conner, F.~Xiao, C.~P.~Landee,
F.~A.~Chaves, S.~Soriano, M.~A.~Novak, T.~P.~Papageorgiou,
A.~D.~Bianchi, T.~Herrmannsd\"{o}rfer, J.~Wosnitza, J.~A.~Schlueter,
Phys. Rev. B {\bf75}, 094421 (2007).

\bibitem{Xiao}F. Xiao, F.~M.~Woodward, C.~P.~Landee, M.~M.~Turnbull,
C.~Mielke, N.~Harrison, T.~Lancaster, S.~J.~Blundell, P.~J.~Baker,
P.~Babkevich, F.~L.~Pratt, Phys. Rev. B {\bf 79}, 134412 (2009).

\bibitem{Cuccoli03}A.~Cuccoli, T.~Roscilde, R.~Vaia, and P.~Verrucchi, Phys. Rev.
B {\bf 68}, 060402(R) (2003).

\bibitem{BKT1}V.~L.~Berezinskii, Sov. Phys. JETP {\bf 32}, 493 (1971).

\bibitem{BKT2}J.~M.~Kosterlitz and D.~J.~Thouless, J.~Phys. C {\bf 6}, 1181 (1973).

%
%

\bibitem{Chernyshev}A.~L.~Chernyshev, Phys. Rev. B, {\bf 72}, 174414 (2005)

\end{thebibliography}
\end{document}